\documentstyle[sprocl]{article}

\begin{document}

\title{BOUND-STATE PROBLEM IN A ONE-DIMENSIONAL CANTOR-LIKE POTENTIAL}
\author{L. D. Almeida, F. Kokubun, D. Hadjimichef}
\address{ Dep. de F\'{\i}sica, Funda\c{c}\~ao Universidade do Rio Grande\\
 Rio Grande, R. S., Brazil}
\maketitle
\abstracts{
One of the best systems for the study of quantum chaos is the atomic nucleus.
A confined particle with 
general boundary conditions can present chaos and the eigenvalue problem can 
exhibit this fact. We study a toy model in which the potential has a
 Cantor-like form. The eigenvalue spectrum presents a {\it Devil's staircase}
 ordering in the semi-classical limit.}

 The spectra of complicated nuclear systems 
have similar statistical properties to those of the spectral of ensembles of 
random matrices. Theoretical work, especially
realistic shell model calculations is helping to establish the domains of
chaos in nuclei. In the chaotic regime a scattered 
particle can be trapped, for some time, by multiple reflection in the 
potential region producing a compound nucleus. The mean field is too regular,
as can be seen by the existence of the shell structure, so chaos must be caused
by the residual interaction. Due to the exclusion principle the role of the
latter increases with the excitation energy \cite{caos2}.

Fractals have received enormous attention as models for the complex and 
irregular \cite{fractal1}. Over the last years, the question of what kind of 
physical phenomena should arise when waves interact with fractal structures
has been addressed by many authors\cite{berry1}. 
Inspired by the problem in nuclear physics we present an exploratory study of 
the eigenvalue problem of a quantum particle trapped in an infinite well 
subject to a Cantor-like potential $V(x)=V_{0} v(x)$ defined in the region 
between the infinite walls.
The dimensionless function $v(x)$ has a minimum value  of $-1$ and a maximum 
value of $+1$ \cite{koonin}. The potential is constructed order by order as
can be seen in Fig. (a). The starting point is a rescaled Schr\"odinger 
equation
\begin{eqnarray}
\left[-\frac{1}{\mu^2} \frac{d^2}{dx^2} + v(x)\right]\psi(x)=
\epsilon\psi(x)
\label{eq_shrod}
\end{eqnarray}
where $\mu=\sqrt{2m\lambda^{2}V_{0}}/ \hbar$ and
$\lambda$ is the scaling length of the coordinate and $\epsilon=E/V_{0}$ is 
the dimensionless energy. One can control the classical nature of the system
by variating the dimensionless measure $\mu$. When one takes 
$\mu\rightarrow\infty$ (which is equivalent to $\hbar\rightarrow 0$) the 
the classical limit is reached. 

We solved numerically the Schr\"odinger equation and obtained the follows results.
For small values of $\mu$
the particle can only be confined in the large structures of the 
potential. But as $\mu$ is increased lower energy values are accessed
and the eigenvalue spectrum is semi-classical. 
In this limit one finds the  result that the particle can be confined
in the small structures of the potential. This effect can be seen in the
following example where we consider $\mu=300$ (Fig (a) ).
The probability density is showed for the eigenvalues 
$\epsilon=-0.31340$ and $\epsilon=-0.31544$.
One sees in Fig. (b) the cumulative density of states where up to $\epsilon=0$ it 
exhibits a {\it Devil's staircase}  behavior. Clustering effects in the spectrum are shown 
in Fig. (c) for a chosen eigenvalue. 

One can conclude from this study that a fractal potential produces highly localized solutions
in the semi-classical limit and clustering effects in the eigenvalue spectrum.
These results indicate that fractal potentials deserve a more detailed attention when
analyzing complex systems.

\vspace{1cm}

{\bf Acknowledgments}

This work was supported by FURG and FAPERGS, Brazilian Financial agency.


\begin{thebibliography}{99999}


\bibitem{caos2} E. Caurier, J. M. G. G\'omez, V. R. Manfredi, L. Salasnich,
{\it Phys. Lett.} B {\bf 365}, 7 (1996).

\bibitem{fractal1} B. B. Mandelbrot, {\sl The Fractal Geometry of Nature}
(Freeman, San Francisco, 1982); {\sl The Fractal Approach to Heterogeneous 
Chemistry: Surfaces, Colloids, Polymers}, ed. by D. Avnir (John Wiley \&
Sons Ltd., Chichester, 1992);{\sl Fractals in Science}, ed. by A. Bunde, S. 
Havlin (Springer, Berlin, 1994).

\bibitem{berry1} M. V. Berry, {\it J. Phys.} A {\bf 12}, 781 (1979);
 J. Goujun, F. Bihua and F. Duan, {\it Chem. Phys. Lett.} {\bf 5}, 
9 (1988); D. Berger, S. Chamaly, M. Perreau, D. Mercier, {\it et al}, 
{\it J. de Phys.} I {\bf 1}, 1433 (1991); 
J. Uozumi, H. Kimura and T. Asakura, {\it J. Mod. Optics} {\bf 38}, 1335 (1991); 
C. Iemmi and S. Ledesma, {\it Optics Comm.} {\bf 112}, 1 (1994); M. V. Berry,
 {\it J. Phys A:Math. Gen.} {\bf 29}, 6617 (1996).


\bibitem{koonin} S. E. Koonin and D. C. Meredith, {\sl Computational Physics},
(Addison-Wesley Publish Company, 1990).

\end{thebibliography}
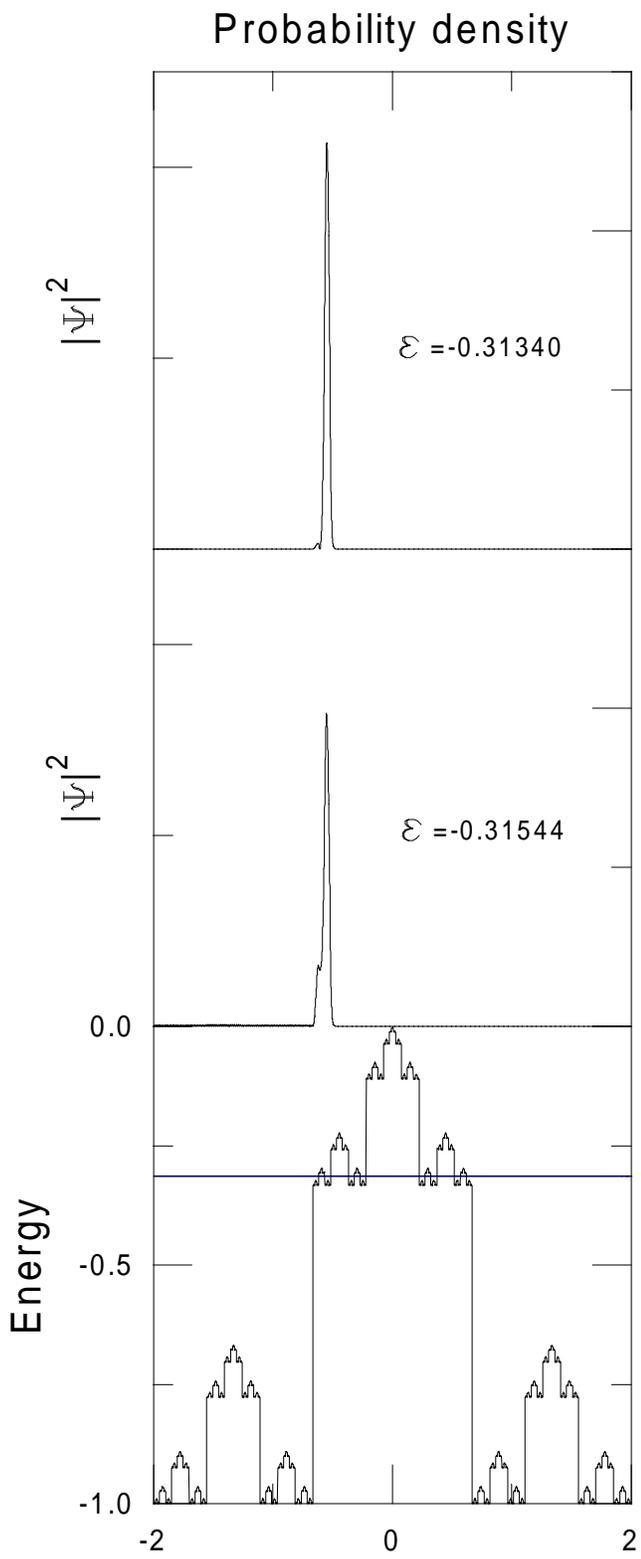
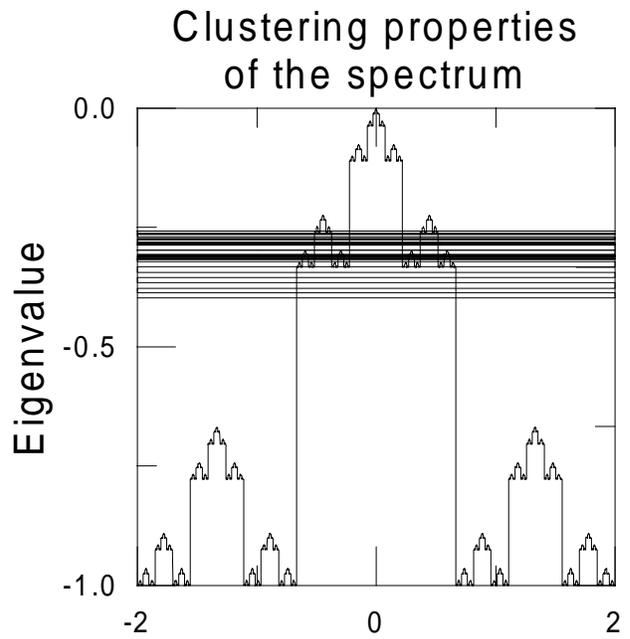
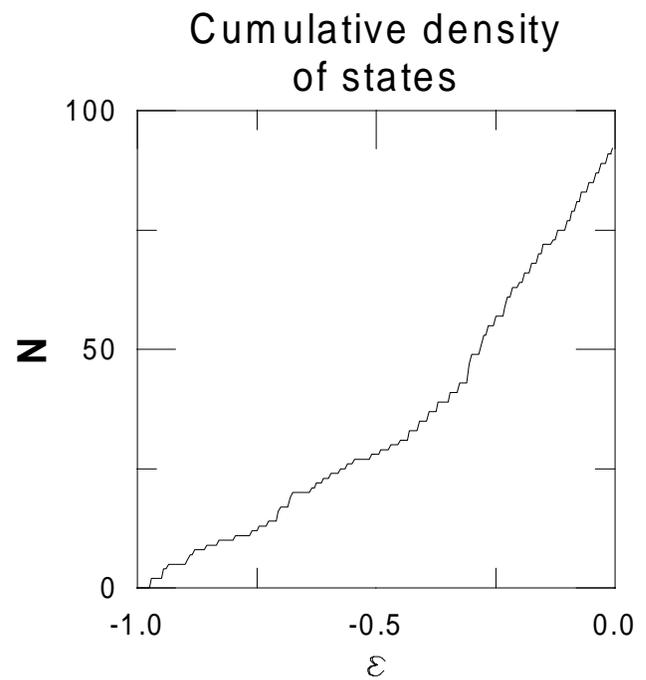
\end{document}